# Using the Standard Linear Ramps of the CMS Superconducting Magnet for Measuring the Magnetic Flux Density in the Steel Flux Return Yoke

Vyacheslav Klyukhin[1,2], Benoit Curé[2], Nicola Amapane[3], Austin Ball[2], Andrea Gaddi[2], Hubert Gerwig[2], Alain Hervé[4], Richard Loveless[4], and Martijn Mulders[2]

[1]Skobeltsyn Institute of Nuclear Physics, Lomonosov Moscow State University, Moscow, RU-119992, Russia
[2]CERN, Geneva 23, CH-1211 Switzerland
[3]INFN Turin and University of Turin, I-10125, Turin, Italy
[4]University of Wisconsin, Madison, WI 53706, USA

The principal difficulty in large magnetic systems having an extensive flux return yoke is to characterize the magnetic flux distribution in the yoke steel blocks. Continuous measurements of the magnetic flux density in the return yoke are not possible and the usual practice uses software modelling of the magnetic system with special three-dimensional (3-D) computer programs. The 10,000-tonne flux return yoke of the Compact Muon Solenoid (CMS) magnet encloses a 3.8 T superconducting solenoid with a 6-m-diameter by 12.5-m-length free bore and consists of five dodecagonal three-layered barrel wheels around the coil and four endcap disks at each end. The yoke steel blocks, up to 620 mm thick, serve as the absorber plates of the muon detection system. A TOSCA 3-D model of the CMS magnet has been developed to describe the magnetic field outside of the solenoid volume, which was measured with a field-mapping machine. To verify the magnetic flux distribution calculated in the yoke steel blocks, direct measurements of the magnetic flux density with 22 flux loops installed in selected regions of the yoke were performed during the CMS magnet test in 2006 when four "fast" discharges of the CMS coil (190 s time-constant) were triggered manually to test the magnet protection system. No fast discharge of the CMS magnet from its operational current of 18.2 kA, which corresponds to a central magnetic flux density of 3.8 T, has been performed that time. For the first time, in this paper we present measurements of the magnetic flux density in the steel blocks of the return yoke based on the several standard linear discharges of the CMS magnet from the operational magnet current of 18.2 kA. To provide these measurements, the voltages induced in the flux loops (with amplitudes of 20–250 mV) have been measured with six 16-bit DAQ modules and integrated offline over time. The results of the measurements during magnet linear ramps performed with a current rate as low as 1–1.5 A/s are presented and discussed.

*Index Terms*—Electromagnetic modeling; flux loops; Hall effect devices; magnetic field measurement; magnetic flux density; measurement techniques; superconducting magnets.

## I. INTRODUCTION

THE PRINCIPAL DIFFICULTY in large magnetic systems having an extensive flux return yoke [1], [2] is to characterize the magnetic flux distribution in the yoke steel blocks. Continuous measurements of the magnetic flux density in the return yoke are not possible and the usual practice uses software modelling of the magnetic system with special three-dimensional (3-D) computer programs [3], [4]. The magnetic flux density in the central part of the Compact Muon Solenoid (CMS) detector [2], one of the large physics detectors located at the Large Hadron Collider (LHC) at CERN (Geneva, Switzerland), was measured with a precision of $7 \times 10^{-4}$ with a field-mapping machine [5] before the solenoidal volume was filled with physics detectors. The magnetic flux everywhere outside of this measured volume was estimated by a 3-D magnetic field model with the program TOSCA [6] from *Cobham CTS Limited*. This model reproduced the magnetic flux density distribution measured with the field-mapping machine inside the CMS coil to within 0.1% [7].

To verify the magnetic flux distribution calculated in the yoke steel blocks, direct measurements of the magnetic flux density in the selected regions of the yoke were performed during the CMS magnet test in 2006 when four "fast" discharges of the CMS coil (190 s time-constant) were triggered manually to test the magnet protection system. These discharges were used to induce voltages with amplitudes of 0.5–4.5 V in 22 flux loops wound around the yoke blocks in special grooves, 30 mm wide and 12–13 mm deep. The loops have 7–10 turns of 45-wire flat ribbon cable and the cross sections of areas enclosed by the flux loops vary from 0.3 to 1.59 m² on the yoke barrel wheels and from 0.5 to 1.12 m² on the yoke endcap disks [8]. An integration technique [9] was developed to reconstruct the average initial magnetic flux density in the cross sections of the steel blocks at full magnet excitation.

The comparisons of the magnetic flux densities measured with the flux loops during the CMS coil fast discharges and the magnetic field values computed with the CMS magnet model are presented elsewhere [8], [10]. At the time those papers were published no fast discharge of the CMS magnet from its operational current of 18.2 kA, which corresponds to a central magnetic flux density of 3.8 T, had been performed.

## II. MATERIALS AND METHODS

### A. Upgrading the Flux Loop Readout System

During the LHC long shutdown of 2013/2014 the readout system of the flux loop voltages was upgraded to replace the 12-bit *USB-1208LS* DAQ modules from *Measurement Computing* with new 16-bit *USB-1608G* modules from the same manufacturer. This replacement allowed measurements of readout voltages with a precision of 0.15 mV compared with the precision of 2.44 mV with the 12-bit modules. The new 16-bit readout gives a resolution of 0.75 % at a typical amplitude of 20 mV. The DAQ modules were attached by USB cables to two network-enabled *AnywhereUSB®/5* hubs connected to the DAQ PC through *3Com® OfficeConnect® Dual Speed Switch 5* and a 100 m optical fiber cable with two *Magnum CS14H-12VDC Convertor Switches*. These modifications permitted a measurement of the magnetic flux density in the steel blocks using standard magnet ramps with a current discharge rate as low as 1–1.5 A/s. To improve the precision of the flux loop measurements, the total areas, $A$, covered by the flux loops have been measured for each individual wire turn position; they vary from 122 to 642 m². This reduced a systematic error arising from the flux loop conductor arrangement to ±3.6 % on average.

The average magnetic flux density components $B_i$ ($i = z, y$) orthogonal to the flux loop cross sections were obtained as $B_i = \Theta/A$, where the magnetic flux $\Theta$ is calculated by integration of the signal voltages over total time of the measurement [9]. The calculations performed with the CMS magnet model have shown that the magnetic flux density is quite uniform in the flux loop cross sections. The flux loop area $A$ is calculated by the following expression: $A = N \times (a + c) \times (b + c) + d$, where $N$ is the total number of the flux loop turns, $a$ and $b$ are the width and height of the 5$^{th}$ turn of the flat ribbon cable, $c$ and $d$ are small constant terms dependent on the number of turns $N$.

### B. CMS Magnet Model

The CMS magnet model used for the magnetic field map preparation and for the comparisons with the measurements was modified to include all the ferromagnetic parts beyond the central magnet yoke as well as the electrical current leads for the solenoid coil as shown in Fig. 1.

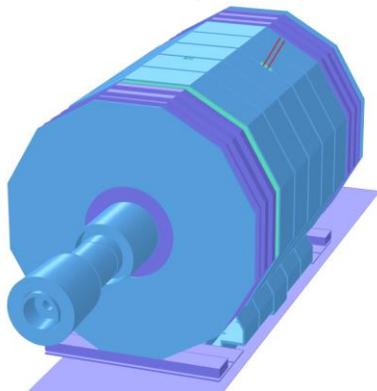

Fig. 1. CMS magnet 3-D model evaluated with the program TOSCA at the operational current of 18.2 kA. Different colors correspond to different steel used in the magnet flux return yoke. Five barrel wheel in the center and four end cap discs at each end of the central 14-m-diameter yoke are shown. The cylinders downstream the central yoke are the forward hadronic calorimeter, collar, beam pipe rotating shielding, and fixed iron nose. The forward part of the model extends to ± 21.89 m in each direction with respect to the coil center. Two electrical current leads supplying the coil with the current of 18.2 kA are visible outside a special chimney.

The coordinate system used in this study corresponds to the CMS reference system where the *X*-axis is aligned in the horizontal plane towards the LHC center on the near side of the CMS detector, the *Y*-axis is aligned upwards, and the *Z*-axis coincides with the superconducting coil axis and has the same direction as the positive axial component of the magnetic flux density.

To perform comparisons with the measurements presented in this study, the magnetic flux density was calculated in the areas where the measuring devices are located on the CMS yoke steel blocks. In addition to the flux loops, the magnetic flux density was also measured with the 3-D Hall sensors installed between the barrel wheels and on the first endcap disk at the axial *Z*-coordinates of 1.273, –1.418, –3.964, –4.079, –6.625, and –7.251 m. The sensors are aligned in rows at the vertical *Y*-coordinates of –3.958, –4.805, –5.66, and –6.685 m [10] on two sides of the magnet yoke: the near side towards the LHC center (positive *X*-coordinates), and the far side opposite to the LHC center (negative *X*-coordinates). In the present analysis, the 3-D Hall sensors installed on the inner surfaces of both CMS yoke nose disks inside the coil were also used. The layout of the flux loops and the 3-D Hall probes has been reported elsewhere [11].

### C. CMS Magnet Model Crosschecks

To crosscheck the model, a comparison of the magnetic flux density from the model to the measurements done with four NMR-probes and four 3-D Hall sensors installed inside the solenoidal volume was made at the operational current of 18.2 kA. Two NMR-probes are located near the coil middle plane at the *Z*-coordinates of ±0.006 m and radii of 2.9148 m; another two probes are installed on the CMS tracker faces at the *Z*-coordinates of –2.835 and +2.831 m and radii of 0.651 m. Four 3-D Hall sensors are located on the CMS tracker faces at the Z-coordinates of –2.899 and +2.895 m and radii of 0.959 m. The averaged precision of the NMR-probe measurements was $(5.2 \pm 1.3) \times 10^{-5}$ T, that of the 3-D Hall sensors was $(3.5 \pm 0.5) \times 10^{-5}$ T. The averaged relative differences between the modelled and measured values of the magnetic flux density were $(-5.6 \pm 1.7) \times 10^{-4}$ at the NMR-probe locations, and $(-2.4 \pm 4.0) \times 10^{-4}$ at the 3-D Hall sensor locations. This close result verifies that the CMS model provides a good description of the magnetic flux distribution inside the solenoidal volume.

## III. RESULTS AND DISCUSSION

The ramping of the CMS magnet occurs only a few times in a year, so collecting data for the flux loop measurements is a challenging procedure. The measurements used for the present comparisons were obtained in three CMS magnet standard discharges from a current of 18.2 kA to zero, carried out in 2015 and 2016 as shown in Fig. 2.

The first discharge, on July 17–18, 2015, was made with a constant current ramp down rate of 1.5 A/s to a current of 1 kA, and after a pause of 42 s, the fast discharge of the magnet was triggered manually to end the ramp down. The measurements of the voltages induced in the flux loops (with maximum amplitudes of 20–250 mV) were integrated over 15061.5 s in the flux loops located on the barrel wheels and over 15561.5 s in the flux loops located on the endcap disks. The preliminary results obtained in this particular magnet ramp down have been published elsewhere [12].

The typical induced voltage in the first magnet ramp down, together with the integrated average magnetic flux density, is shown in Fig. 3. The rapid maximum and minimum voltage at 11445 s corresponds to the pause in the ramp down at a current of 1 kA, and the following transition from the standard ramp down to the fast discharge of the magnet on the external resistor.

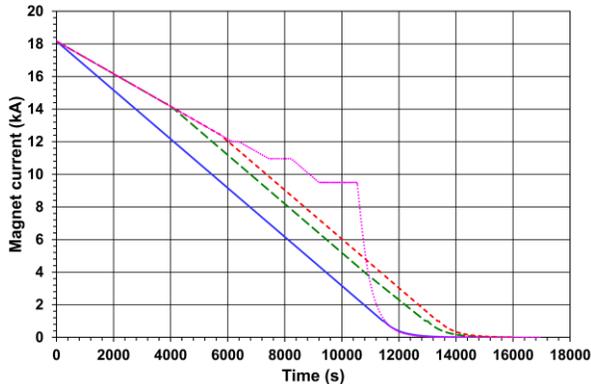

Fig. 2. CMS magnet current discharges from 18.2 kA to zero made on July 17–18, 2015 (*smooth line*), September 21–22, 2015 (*dashed line*), September 10, 2016 (*small dashed line*), and November 30, 2017 (*dotted line*).

The second magnet discharge, on September 21–22, 2015, was performed with two constant ramp down rates: 1 A/s to a current of 14.34 kA (3 T central magnetic flux density), and 1.5 A/s to a current of 1 kA.

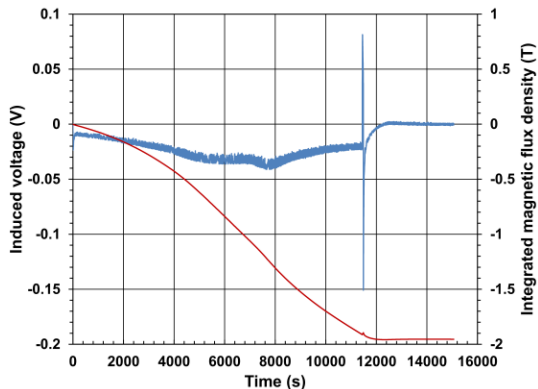

Fig. 3. The induced voltage (*left scale*, *noisy curve*) and the integrated average magnetic flux density (*right scale*, *smooth curve*) in the cross section at $Z = -2.691$ m of the first layer block of the barrel wheel adjacent to the central wheel.

The third magnet discharge, on September 10, 2016, was similar, but the current at which the rate transitioned from 1 A/s to 1.5 A/s was 12.48 kA. Changing the current rates was required by the cryogenic system operational conditions. In both these magnet ramp downs the fast discharges were triggered from a current of 1 kA, and the offline integration of the induced voltages was performed over 17000 s.

In Figs. 4, 5, and 6, the measured values (filled markers) of the magnetic flux density vs. $Z$- and $Y$-coordinates are displayed and compared with the field values computed by the CMS model (open markers) at the operational current of 18.2 kA. The lines shown in Figs. 4 and 5 represent the magnetic flux densities modelled along the lines across the $XY$-coordinates of the Hall sensors those are from 0.155 to 1.325 m away of the flux loop central $XY$-coordinates.

These comparisons give the following differences between the modelled and measured values of the magnetic flux densities in the flux loop cross sections: 4.3 ± 7.0 % in the barrel wheels and –0.6 ± 3.0 % in the endcap disks.

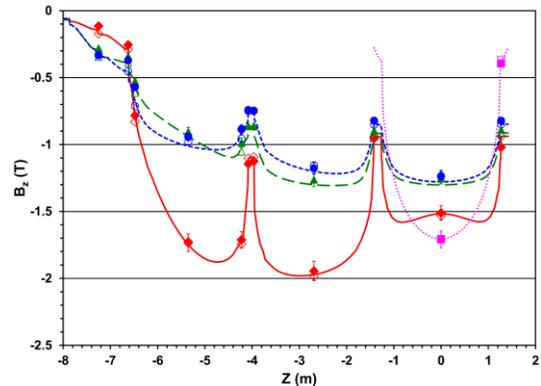

Fig. 4. Axial magnetic flux density measured (*filled markers*) and modelled (*open markers*) in the tail catcher (*squares*) and the first (*diamonds*), second (*triangles*), and third (*circles*) barrel layers vs. the $Z$-coordinate at the near side of the yoke and the $Y$-coordinates of −3.958 m (*dotted line*), −4.805 m (*solid line*), −5.66 m (*dashed line*), and −6.685 m (*small dashed line*).

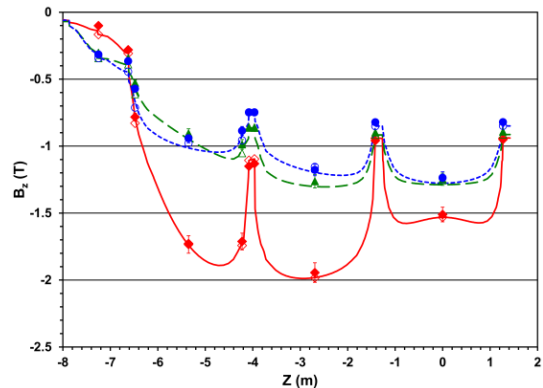

Fig. 5. Axial magnetic flux density measured (*filled markers*) and modelled (*open markers*) in the first (*diamonds*), second (*triangles*), and third (*circles*) barrel layers vs. the $Z$-coordinate at the far side of the yoke and the $Y$-coordinates of −4.805 m (*solid line*), −5.66 m (*dashed line*), and −6.685 m (*small dashed line*).

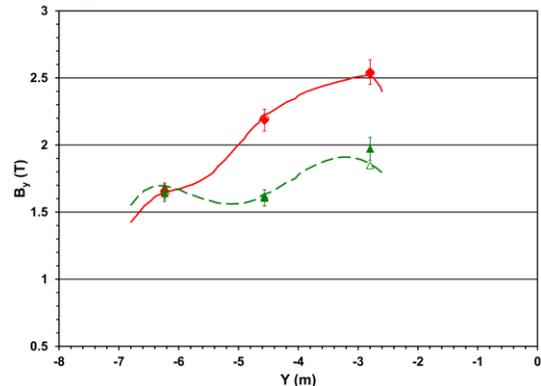

Fig. 6. Radial magnetic flux density measured (*filled markers*) and modelled (*open markers*) in the first (*diamonds*) and second (*triangles*) endcap disks vs. the $Y$-coordinate. The lines represent the calculated values along the lines across the centers of the flux loops.

The errors of the magnetic flux density measured with the flux loops include the standard deviation in the set of three measurements (11 ± 10 mT or 0.86 ± 0.69 % on average) and a systematic error of ±3.6 % arising from the flux loop conductor arrangement. The difference between the modelled and measured magnetic flux density in the 3-D Hall sensor locations is 3 ± 7 %. The error bars of the 3-D Hall sensor measurements are ± (0.02 ± 0.01) mT.

After the latest measurements, comparisons of the calculated values of the magnetic flux density in the yoke steel blocks and the measured values obtained in 2006 with the fast discharges of the magnet [10] have been revised. The forth magnet discharge of November 30, 2017 presented in Fig. 2 was used to exclude the eddy current contributions from the induced voltages of the 2006 measurements. The revised differences between the calculations done with the latest CMS magnet model and the 2006 measurements are as follows: 1.1 ± 7.6 % in the barrel wheels and −0.3 ± 2.0 % in the endcap disks at a maximum current of 17.55 kA; 0.5 ± 7.2 % in the barrel wheels and 0.9 ± 2.2 % in the endcap disks at a maximum current of 19.14 kA. This is compatible with the latest measurements.

Flux loop measurements of the magnetic flux density in the steel blocks of the CMS magnet yoke during a fast discharge are extremely difficult. In the 2006 measurements, the detector was not in the full configuration [8], [10]. The signal voltages had amplitudes of 0.5 – 4.5 V, but were exposed to eddy currents of at least 1 – 2.5 %, as was calculated [13]. Basing on the magnet fast discharge made on November 30, 2017 from the current of 9.5 kA (2 T central magnetic flux density) the eddy current contributions to the 2006 measurements are estimated to be 5.3 ± 4.9 % in the barrel wheels, and 5.5 ± 3.4 % in the endcap disks.

An attempt to reduce the eddy current contribution with an integration of the voltages induced in the flux loops during the standard magnet ramps has been made before upgrading the flux loop readout system, but gave very large errors due to the reading of very small voltages with the previous 12-bit DAQ modules. An upgrade of the readout electronics and a revision of the areas enclosed by the flux loops made it possible to use the standard ramps of the CMS magnet to avoid the large eddy current contribution. Stability of these measurements confirms the correctness of the CMS magnetic field description calculated with the CMS magnet model in TOSCA.

## IV. Conclusion

For the first time, reliable measurements of the magnetic flux density in the steel blocks of the CMS magnet flux return yoke have been made using the flux loop technique and standard magnet discharges from an operational current of 18.2 kA to zero with a current ramp down rate of 1–1.5 A/s. The precision of the measurements is compatible with that of results obtained in 2006, which used the fast discharges of the magnet from similar current values. These new measurements confirm that the new DAQ system is able to monitor the magnetic flux density in the CMS yoke during any standard magnet ramp as well as prove that the latest CMS magnet model provides us with reliable magnetic flux density values across all the CMS detector volume.